\documentclass[a4paper,twocolumn,showpacs,amsmath, amssymb,nofootinbib,aps,floatfix]{revtex4}
\usepackage{epsfig}
\input psfig.sty %{epsf}

\usepackage{bm}									% Bold math
\usepackage{dcolumn}								% Align table columns on decimal point
\usepackage{graphicx}								% Include figure files
\usepackage{hyperref}

\newcommand{\no}{\mbox{$n_{e_o}$}}

\begin{document}

\title{Model-independent constraints on the cosmic opacity}

\author{R. F. L. Holanda$^{2}$\footnote{E-mail: holanda@uepb.edu.br}}

\author{ J. C. Carvalho$^1$\footnote{E-mail: carvalho@dfte.ufrn.br}}

\author{J. S. Alcaniz$^1$\footnote{E-mail: alcaniz@on.br}}

\address{$^1$Departamento de Astronomia, Observat\'orio Nacional, 20921-400, Rio de Janeiro - RJ, Brasil}

\address{$^2$Departamento de F\'{\i}sica, Universidade Estadual da Para\'{\i}ba, 58429-500, Campina Grande - PB, Brasil}

\date{\today}

\begin{abstract}
We use current measurements of the expansion rate $H(z)$ and cosmic background radiation bounds on the spatial curvature of the Universe to impose cosmological model-independent constraints on cosmic opacity. To perform our analyses, we compare opacity-free distance modulus from $H(z)$ data with those from two supernovae Ia compilations:  the Union2.1 plus the most distant  spectroscopically confirmed SNe Ia  (SNe Ia SCP-0401 $z=1.713$) and two Sloan Digital Sky Survey (SDSS) subsamples.  The influence of different SNe Ia light-curve fitters (SALT2 and MLCS2K2) on the results is also verified. We find that a completely transparent universe  is in agreement with the largest sample in our analysis (Union 2.1 plus SNe Ia SCP-0401). For SDSS sample a such universe it is compatible at  $< 1.5\sigma$ level  regardless the SNe Ia light-curve fitting used.

\end{abstract}

%\pacs{98.80.-k, 98.80.Es, 98.65.Cw}

\maketitle

\section{Introduction}

Type Ia supernovae (SNe Ia) observations provide the most direct evidence for the current cosmic acceleration. In the context of Einstein's general theory of relativity, this result implies either the existence of some sort of dark energy, constant or varying slowly with time and space (see Caldwell and Kamionkowski (2010) and Li {\it et al.} 2011 for recent reviews), or that the matter content of the universe is subject to dissipative processes (Lima \& Alcaniz, 1999; Chimento {\it et al.} 2003).

However, there are still some possible loopholes in current SNe Ia observations and alternatives mechanisms contributing to the acceleration evidence or even mimicking the dark energy behavior have been proposed. Examples are possible evolutionary effects in SNe Ia events (Drell, Loredo \&  Wasserman 2000; Combes 2004), local Hubble bubble (Zehavi et al. 1998; Conley et al. 2007), modified gravity (Ishak, Upadhye \& Spergel 2006; Kunz \& Sapone 2007; Bertschinger \& Zukin 2008),  unclustetered sources of light attenuation (Aguirre 1999; Rowan-Robinson 2002; Goobar, Bergstrom \&  Mortsell 2002) and the existence of Axion-Like-Particles (ALPs), arising in a wide range of well-motivated high-energy physics scenarios, and that could lead to the dimming of SNe Ia brightness (Avgoustidis et al. 2009, 2010).

On the other hand, several authors have recently discussed how the so-called cosmic distance duality (CDD)
\begin{equation}
 \frac{D_{\scriptstyle L}}{D_{\scriptstyle A}}{(1+z)}^{-2}=1\;,
 \label{rec}
\end{equation}
relating the luminosity distance ($D_{\scriptstyle L}$) to the angular diameter distance ($D_{\scriptstyle A}$) of a given source can be used to verify the existence of exotic physics as well as the presence of systematic errors in SNe Ia observations (Basset \& Kunz 2004; Uzan, Aghanim \& Mellier 2004; Holanda, Gon\c{c}alves \& Alcaniz 2012).  As it is well known, the CDD relation is closely connected with the Etherington's reciprocity law (Etherington 1933, Ellis 2007), being valid for {all} cosmological models based on Riemannian geometry and requiring only that source and observer are connected by null geodesics in a Riemannian space-time and that the number of photons is conserved (Ellis 2007).

The implementation of the CDD tests mentioned above follows different routes. Basset \& Kunz (2004), for instance, used SNe Ia data as measurements of $D_{\scriptstyle L}$ and estimates of $D_{\scriptstyle A}$ from FRIIb radio galaxies (Daly \& Djorgovski 2003) and ultra compact radio sources (Gurvitz 1994; Lima \& Alcaniz 2002; Alcaniz 2002) to test possible new physics with basis on a generalization of the CDD relation. Uzan, Aghanim \& Mellier (2004) argued that the Sunyaev-Zel'dovich effect plus X-ray techniques for measuring $D_{\scriptstyle A}$ from galaxy clusters is strongly dependent on the validity of this relation.  By assuming a deformed CDD relation,  $D_L(1+z)^{-2}/D_A=\eta$, in a $\Lambda$CDM background they found that the value $\eta=1$ is only marginally consistent with the galaxy clusters data used in the analysis (for the role of the cluster geometry in this test, see Holanda, Lima \& Ribeiro, 2010; 2012). More recently, a cosmological model-independent test involving only measurements of 
the gas mass fraction of galaxy clusters from Sunyaev-Zeldovich and X-ray surface brightness observations was discussed by Holanda, Gon\c{c}alves \& Alcaniz (2012). From this analysis, no significant violation of the CDD relation was found, with the value $\eta = 1$ lying in the $1\sigma$ interval (for other CDD analyses, we refer the reader to Holanda, Lima \& Ribeiro 2010, 2011, 2012; Nair, Jhingan \& Jain 2011; Li, Wu \& Yu 2011; Meng et al. 2011; Gon\c{c}alves, Holanda \& Alcaniz, 2011; Lima, Cunha \& Zanchin, 2011).

In a different approach, Avgoustidis {et al.} (2010) explored consistency among different distance measurements by considering a possible violation of cosmological photon conservation as the only source of CDD violation.  The authors assumed a flat $\Lambda$CDM model and the distance parameterization, $D_L=D_A(1+z)^{2+\epsilon}$, to place bounds on the CDD parameter $\epsilon$ by combining the SNe Ia Union compilation (Kowalski et al. 2008) with the latest  measurements of the Hubble expansion in the range $0 <  z < 2$ (Stern et al. 2010). The basic idea behind the test proposed is that, while SNe Ia observations are affected by at least four different sources of opacity, namely, the Milky Way, the hosting galaxy,  intervening galaxies, and the Intergalactic Medium, the current $H(z)$ measurements  are obtained from ages estimates of old passively evolving galaxies, which relies only on  the detailed shape of the galaxy spectra, not on the galaxy luminosity. Therefore, differently from $D_{\scriptstyle L}$ 
measurements from SNe Ia, $H(z)$ observations are not affected by cosmic opacity $\tau(z)$ since this quantity is assumed to be not strongly wavelength dependent on the optical band (see Avgoustidis et al. 2009 and references therein for more details). { Using} a direct relation between the CDD parameter $\epsilon$ and $\tau(z)$,  Avgoustidis et al. (2010) found  $\epsilon=-0.04_{-0.07}^{+0.08}$ (2$\sigma$).

{ In this paper, we discuss a cosmological model-independent version of the  CDD test proposed by Avgoustidis {\it{et al.}} (2010). Our approach differs from the one described above in that opacity-free estimates of $D_{\scriptstyle L}$ are obtained from a numerical integration of current $H(z)$ data points and not in the context of a given cosmological model. Constraints on $\epsilon$ are derived by comparing these estimates with SNe Ia observations from the Union 2.1 sample (Suzuki et. al 2012) and two compilations of the nearby + SDSS + ESSENCE + SNLS + Hubble Space Telescope set (Kessler et al. 2009) -- with which we discuss the influence of the different SNe Ia light-curve fitters (SALT2 and MLCS2K2) in the results. It is important to stress that we added at Union 2.1 sample  the most distant  spectroscopically confirmed SNe Ia  (SNe Ia SCP-0401 $z=1.713$)(Rubin et al. 2013). From the analyses performed, we find that the larger sample (Union 2.1 plus the SNe Ia SCP-0401) is compatible with a flat and 
transparent universe at $1\sigma$ level. Moreover, we find that the results depend weakly upon the SNe Ia light-curve fitting, with both SALT2 and MLCS2K2 SNe Ia samples being compatible with a transparent universe at $< 1.5\sigma$ level. }

\begin{figure}
\centerline{\psfig{figure=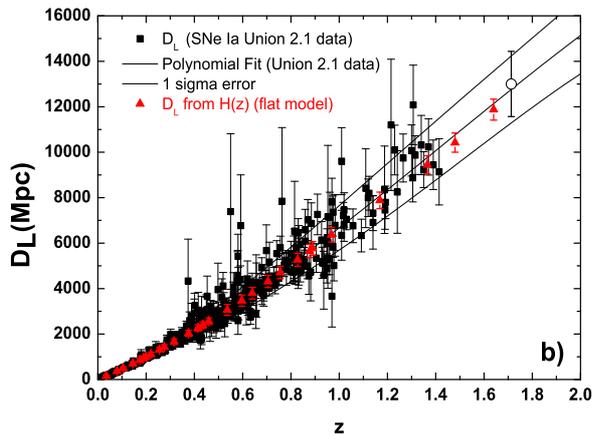,width=3.2truein,height=2.4truein}\hskip 0.1in}
\caption{$D_{\scriptstyle L}(z)$ obtained from measurements of the expansion rate (filled red circle). For the sake of comparison,  we also plot $D_{\scriptstyle L}$ measurements extracted from the Union2.1 SNe Ia sample (filled black squares). The curves stand for the second degree polynomial fit of the opacity-free $D_{\scriptstyle L}$ points from SNe Ia data and the corresponding 1$\sigma$ error. The open circle corresponds to the most distant ($z=1.713$) spectroscopically confirmed SNe Ia.} \label{Fig1}
\end{figure}

\begin{figure*}
\label{Fig}
\centerline{
\includegraphics[width=2.3truein,height=2.4truein]{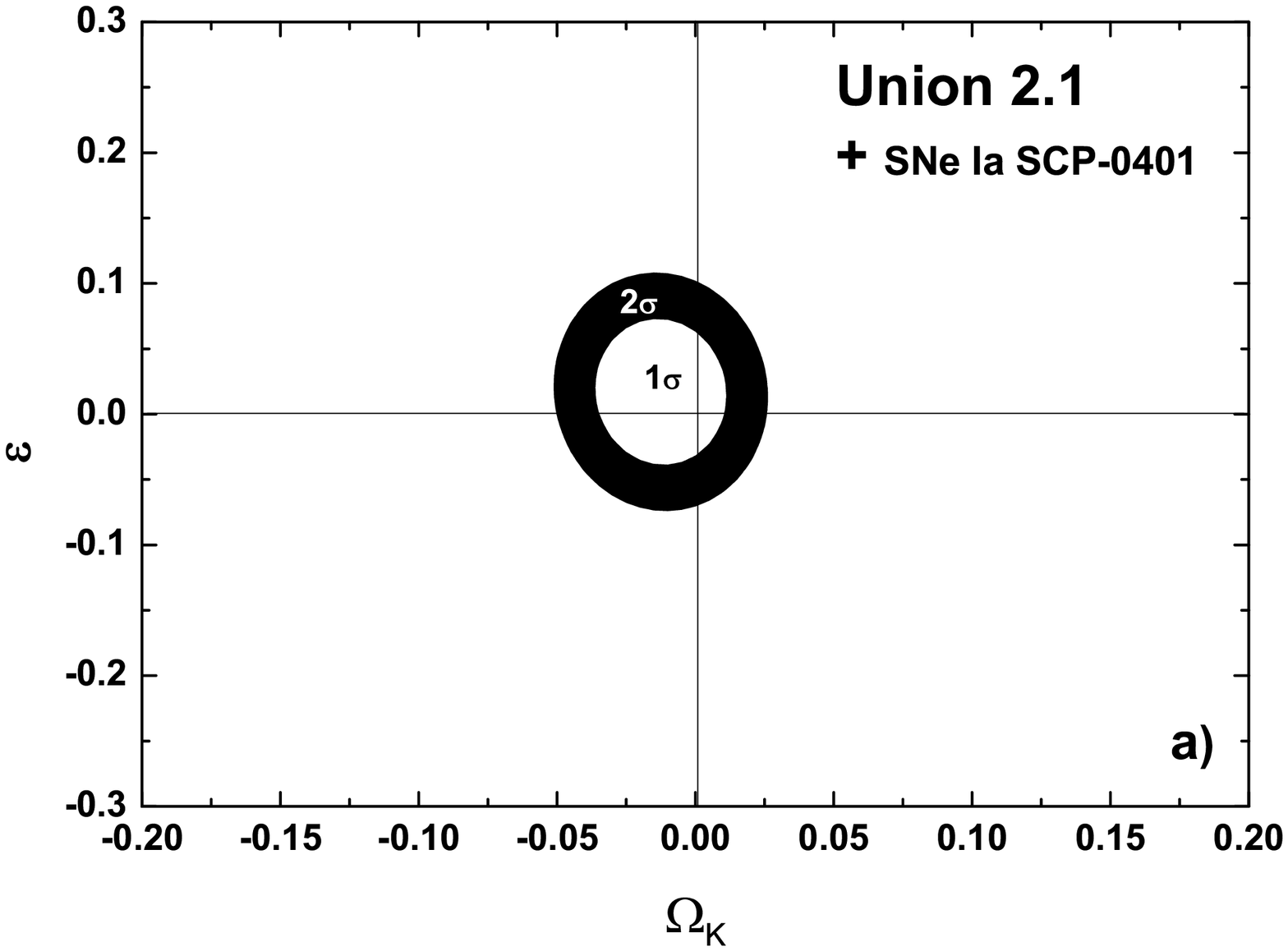}
\includegraphics[width=2.3truein,height=2.4truein]{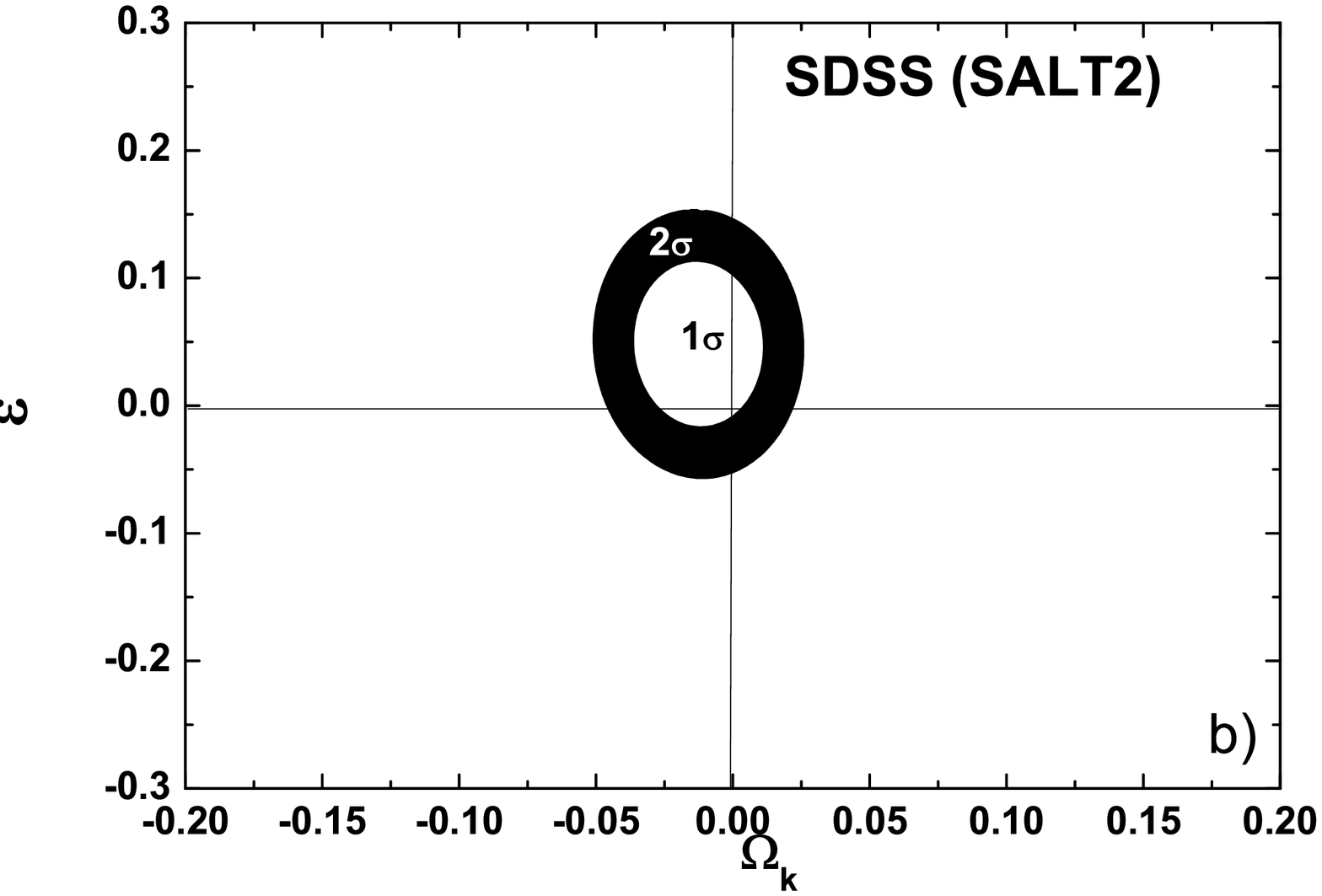}
\includegraphics[width=2.3truein,height=2.4truein]{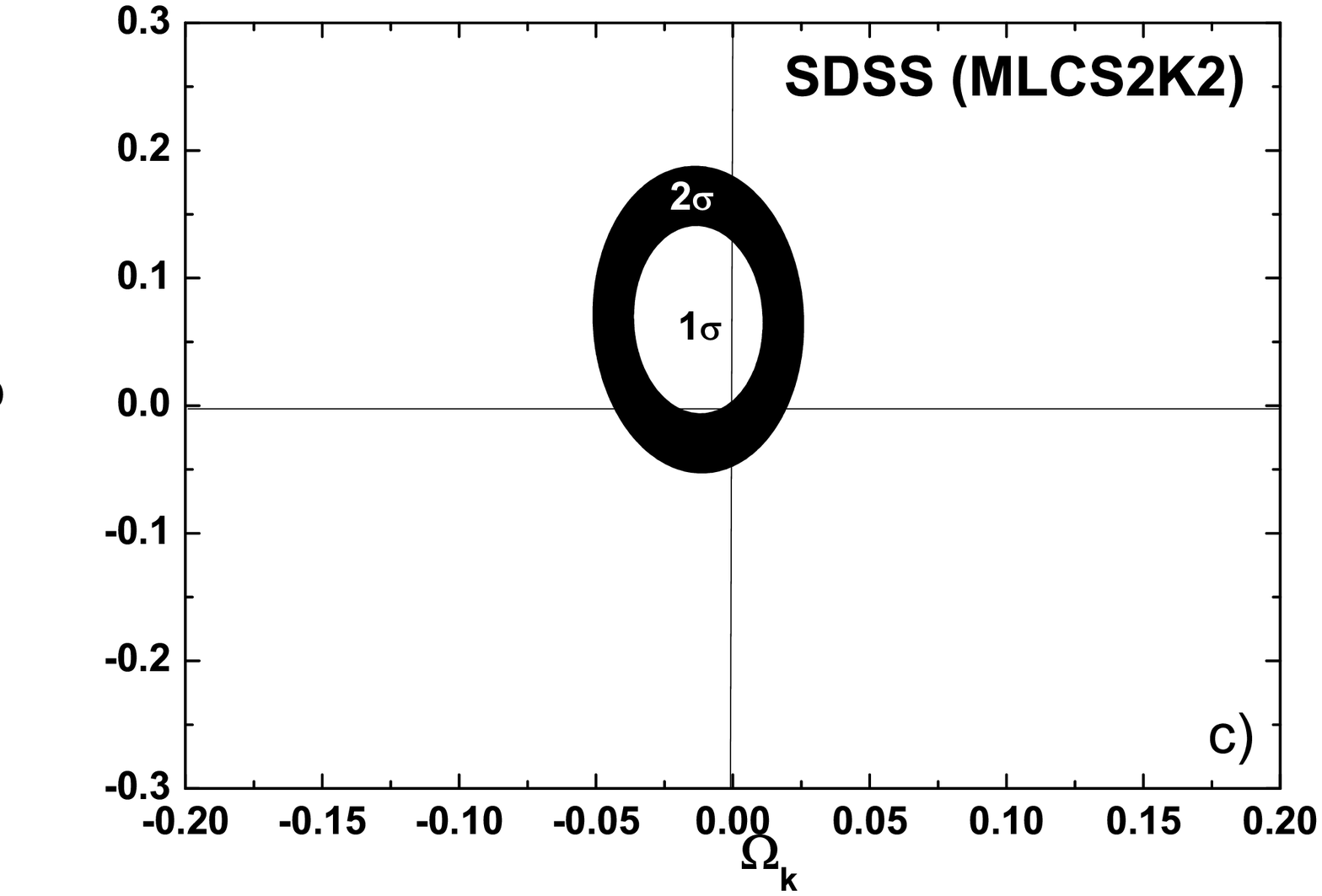}
\hskip 0.1in}
\caption{{ Confidence contours on the plane $\Omega_k - \epsilon$ for the three SNe Ia sub-samples discussed in the text. Contours are drawn for $1\sigma$ and $2\sigma$. The point where the two solid lines crosses stand for a perfect transparent ($\epsilon = 0$) and flat universe  ($\Omega_k = 0$). From these panels, it is clear that the results are weakly dependent upon the SNe Ia light-curve fitting.}}
\end{figure*}

\section{Luminosity distance from $H(z)$ measurements}

Current measurements of the expansion rate at $ z \neq 0 $ are obtained by calculating the derivative of cosmic time with respect to redshift, i.e.,
$ H(z) \simeq - \frac{1}{(1+z)} \frac{\Delta z}{\Delta t}$\footnote{Direct measurements of $H(z)$ at different redshifts is also possible through measurements of the line-of-sight or radial component of baryonic acoustic oscillations (BAO) from large redshift surveys with redshift precision of the order of $0.003(1 + z)$ -- see, e.g., Benitez {\it{et al.}} 2009.}. This method was first presented by Jimenez and Loeb (2002) and consists in measuring the age difference between two red galaxies at different redshifts in order to obtain the rate $ \Delta z / \Delta t $. {In this work, we use the largest $H(z)$ data sample to date which consists of 28 points (Simon et al. 2005, Gaztnaga et al. 2009, Stern et al. 2010 \& Moresco et al. 2012, Blake et al. 2012, Zhang et al. 2012).} 

In order to transform these $H(z)$ measurements into distance estimates, we solve numerically the comoving distance integral for non-uniformly spaced data
\begin{equation}
\label{eq2}
D_C =c \int_0^z{dz^\prime \over H(z^\prime)}\approx {c\over 2}\sum_{i=1}^{N} (z_{i+1}-z_i)\left[ {1\over H(z_{i+1)}}+{1\over H(z_i)} \right],
\end{equation}
using a simple trapezoidal rule. Since the error on $z$ measurements is negligible, we only take into account the uncertainty on the values of $H(z)$.
As one may check, by using standard error propagation techniques, the error associated to the $i^{th}$ bin is given by
\begin{equation}
s_i={c\over 2}(z_{i+1}-z_i)\left({\sigma_{H_{i+1}}^2\over H_{i+1}^4} + {\sigma_{H_{{i}}}^2\over H_{i}^4}\right)^{1/2}\;,
\end{equation}
so that the error of the integral (\ref{eq2}) in the interval $z=0$ -- $z_{n}$ is $\sigma^2_n = \sum_{i=1}^n s_i$.  In order to obtain robust results from our analysis, we added to our $H(z)$ sample, the  value of the expansion rate today $H_0=73.8 \pm 2.4$ km/s/Mpc, as obtained by Riess et al. (2011). The $D_{\scriptstyle L}$ points (see Eq. \ref{dl}) obtained from $H(z)$ observations are shown in Fig. 1 assuming $\Omega_k=0$.  For the sake of comparison, we also plot $D_{\scriptstyle L}$ measurements extracted from the Union 2.1 SNe Ia sample + SNe Ia SCP-0401.

\section{Constraints on cosmic opacity}

\subsection{Methodology}

As argumented by Avgoustidis {\it{et al.}} (2010), if there were a source of photon  absorption affecting the universe transparency, the distance modulus derived from Supernovae would be systematically affected. In particular, any effect that reduces the number of photons would dim the SNe Ia brightness and increases $D_L$. Thus, if $\tau(z)$ denotes the opacity  between an observer at $z=0$ and a source at $z$ due to, e.g., extinction, the flux received from the source would be attenuated by a factor $e^{-\tau(z)}$ and thus the observed luminosity distance ($D_{L, obs}$) is related to the true luminosity distance ($D_{L, true}$) by
 \begin{equation}
 D_{L, obs}^2=D_{L,true}^2 e^{\tau(z)} \, .
 \end{equation}
 Therefore, the \emph{observed} distance modulus is given by (Chen and Kantowski, 2009a; 2009b)
 \begin{equation}
 \label{distancemod}
 m_{obs}(z)=m_{true}(z)+2.5[\log e] \tau(z) \, .
 \end{equation}
In our analyses, measurements of $m_{obs}$ are taken from the SNe Ia Union2.1 and SDSS compilations (see Section 3.2). Note that, differently from the Avgoustidis {\it{et al.}} (2010) analysis, where a flat $\Lambda$CDM model was assumed, we compare $D_{L, obs}$ estimates from SNe Ia data to opacity-free luminosity distance $D_{L,true}$ inferred directly from the $H(z)$ measurements, as described earlier. Note also that, since the SNe Ia and $H(z)$ observations are performed at different $z$, we calculate $D_{L,obs}$ at each SNe Ia redshift from a second degree polynomial fit of the SNe Ia data points shown in Fig. 1. It is worth mentioning that, if instead of fitting $D_L$ from SNe Ia data to compare with $D_L$ points from the expansion rate measurements one does the other way around, the error on the opacity parameter $\epsilon$ becomes underestimated by a factor of $\simeq 2$. This is basically due to the difference in the size of SNe Ia and $H(z)$ samples.

\begin{figure*}
\label{Fig}
\centerline{
\includegraphics[width=2.8truein,height=2.4truein]{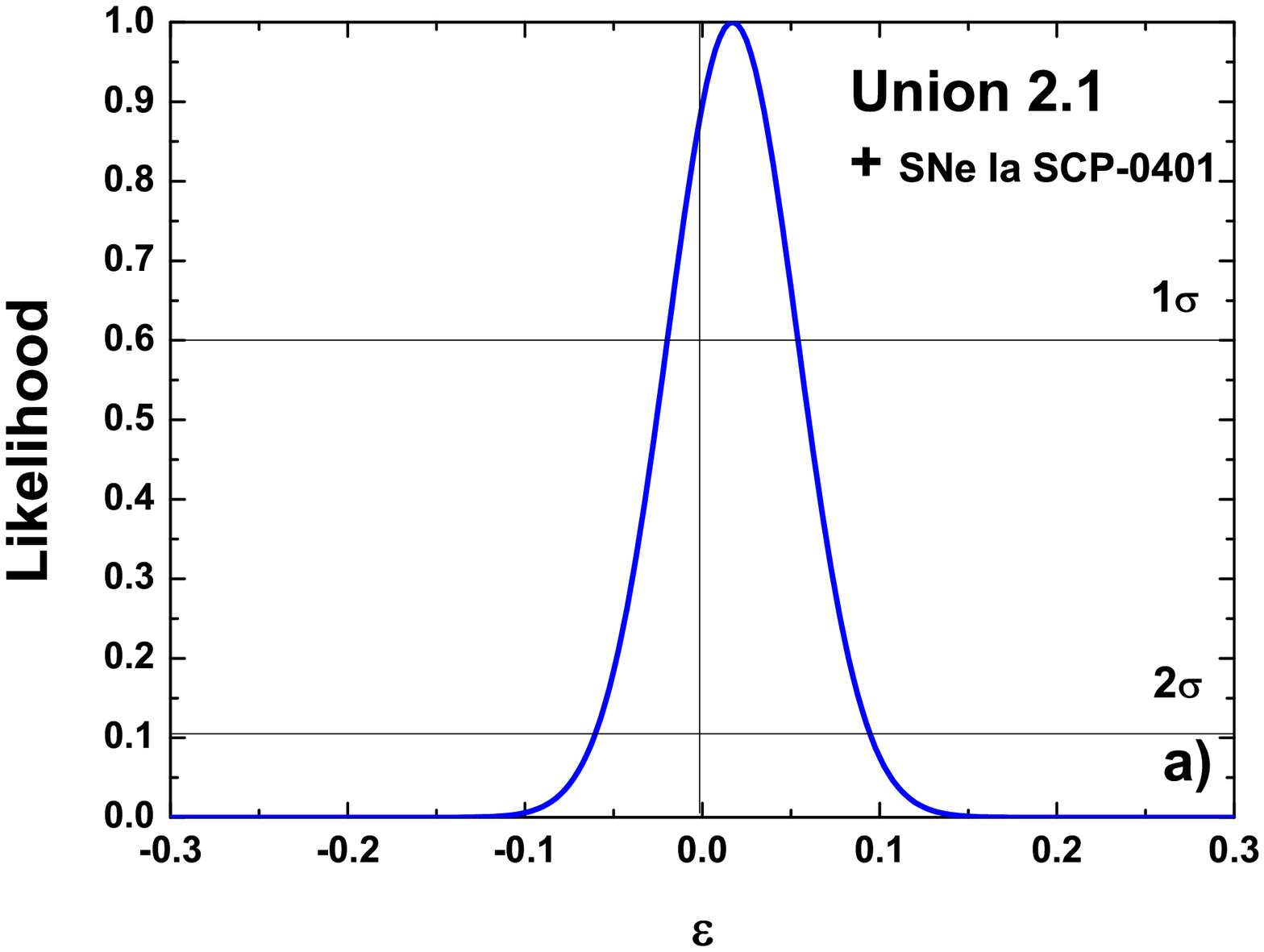}
\includegraphics[width=2.8truein,height=2.4truein]{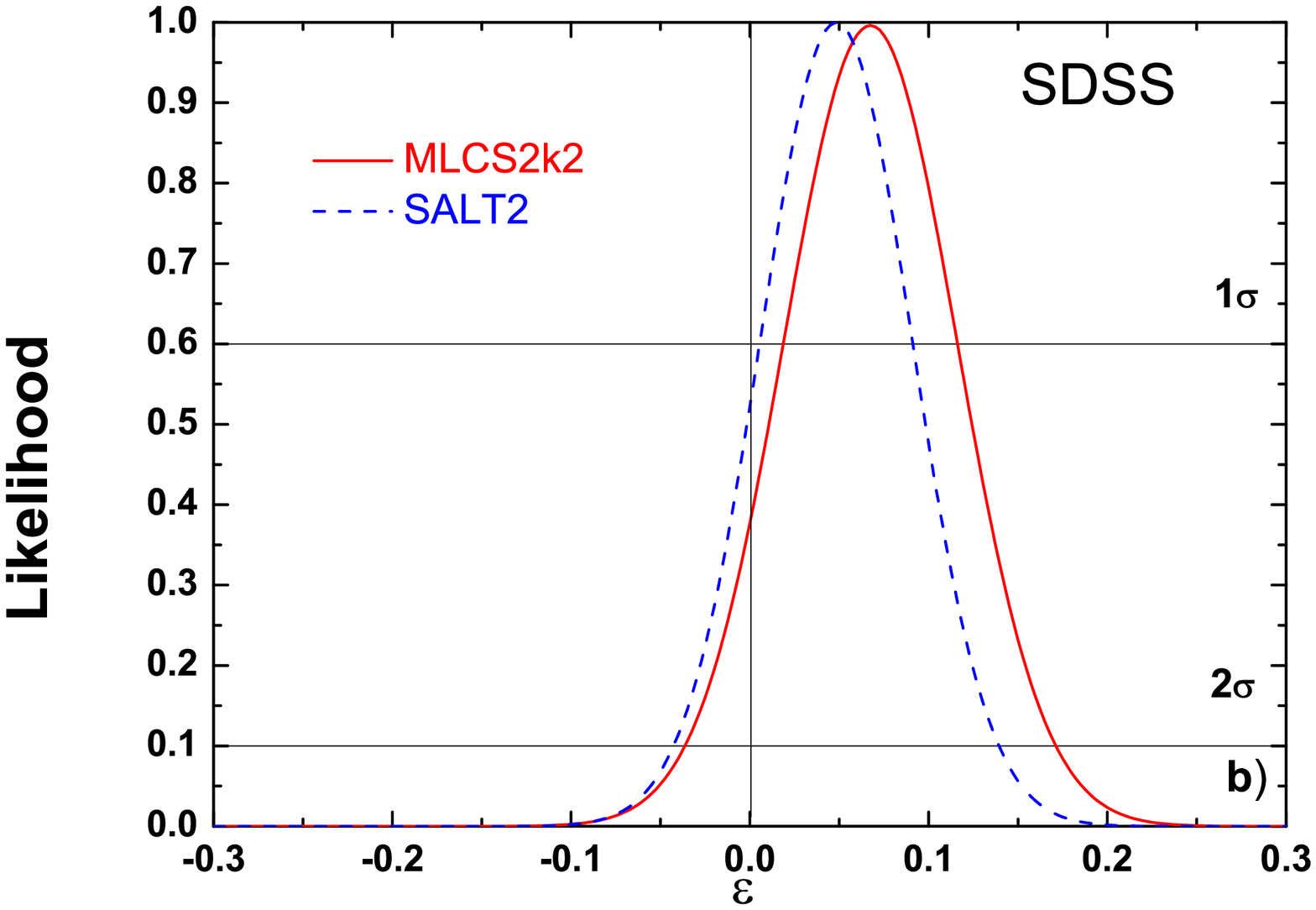}
\hskip 0.1in}
\caption{{Likelihood function for $\epsilon$ after marginalizing over the curvature and Hubble parameters. Note that the largest SNe Ia sample ir our analysis (Union 2.1 plus SNe Ia SCP-0401)  is in full agreement with a transparent universe ($\epsilon = 0$).}}
\end{figure*}

In order to proceed further, we must assume an appropriated redshift dependence for $\tau(z)$. Here, we follow Avgoustidis {\it{et al.}} (2009; 2010) and consider
\begin{equation}
\tau(z)=2\epsilon z\;.
\end{equation}
As shown in the above reference, this linear expression can be derived from the usual CDD parameterization $D_L=D_A(1+z)^{2+\epsilon}$ for small values of $\epsilon$ and $z \leq 1$, where $\epsilon$ quantifies departures from transparency. To discuss our results in a more general setting, we also consider Fridmann-Robertson-Walker geometries with arbitrary curvature and use in our analysis the current WMAP bound on the curvature parameter, i.e., $\Omega_k=0.0125\pm 0.0155$ ($2\sigma$) (Komatsu et al. 2011).

\subsection{Data sets and analysis}

 As mentioned earlier, we use two SNe Ia data sets in our analyses. The Union 2.1 sample is an update of the original Union compilation (Amanullah et al. 2010) that comprises 580 data points including recent large samples from other surveys and uses SALT2 for SNe Ia light-curve fitting.{  Recently, the supernova cosmology project reported the discovery of the most distant SNe Ia (Rubin et al. 2013): SNe Ia SCP-0401. The SNe Ia SCP-0401 has $z=1.713$ and a distance modulus  of $45.57 \pm 0.24$ (statistical errors). In our analysis we have added this SNe Ia to Union 2.1 sample
(see fig. 1b)}. The second sample is the nearby + SDSS + ESSENCE + SNLS + Hubble Space Telescope set of 288 SNe Ia (throughout this paper we refer to this set as SDSS compilation) that uses both SALT2 (Guy et al.\ 2007) and MLCS2K2 (Jha et al.\ 2007) light-curve fitters~\footnote{MLCS2K2 calibration uses a nearby training set of SNe Ia assuming a close to linear Hubble law, while SALT2 uses the whole data set to calibrate empirical light curve parameters, and a cosmological model must be assumed in this method. Typically a $\Lambda$CDM or a $\omega$CDM model is assumed. Consequently, the SNe Ia distance moduli obtained with SALT2 fitter retain a degree of model dependence (Bengochea 2011).} and is distributed in redshift interval $0.02 \leq  z \leq 1.55$. In order to 
verify the effect of the light-curve fitting on the results, we consider both SDSS sub-samples in our analyses. { For this case, we neglected two $H(z)$ measurements: $H(z=1.53)=140 \pm 14$ and $H(z=1.75)=202 \pm 40$.} 

We estimated the best-fit to the set of parameters ${\mathbf{P}} \equiv (\Omega_k,\epsilon, H_0)$, by evaluating the likelihood distribution function, ${\cal{L}} \propto e^{-\chi^{2}/2}$, with
\begin{eqnarray}
\chi^{2} & = & \sum_{z}\frac{(m_{obs}(z) - m_{true}(z)-2.1715\epsilon z)^2}{\sigma^2_{m(obs)}+ \sigma^2_{m(true)} } \\ & & \nonumber + \frac{(\Omega_k+0.0125)^{2}}{0.0155^{2}}
 + \frac{(H_0 - 73.8)^{2}}{2.4^{2}},                                                                
\end{eqnarray}
where $\sigma^2_{m(true)}$ and $\sigma^2_{m(obs)}$  are the errors associated to distance modulus from $H(z)$ measurements and distance modulus from SNe Ia  ({ without systematic errors}), respectively. $m_{true}$ is obtained via $m_{true}=5\log_{10} D_{L,true} + 25$, while $D_{L,true}$ is given by one the following forms (Hogg 2000):
\begin{equation} 
\label{dl}
\frac{D_{L,true}}{(1+z)} = \left\{
\begin{array}{ll}
\,\frac{D_{\rm H}}{\sqrt{|\Omega_k|}}\,\sinh\left[\sqrt{\Omega_k}\,D_{\rm C}/D_{\rm H}\right] & {\rm for}~\Omega_k>0 \\
\,D_{\rm C} & {\rm for}~\Omega_k=0 \\
\,\frac{D_{\rm H}}{\sqrt{|\Omega_k|}}\,\sin\left[\sqrt{|\Omega_k|}\,D_{\rm C}/D_{\rm H}\right] & {\rm for}~\Omega_k<0
\end{array}
\right.
\end{equation}
where $D_H=cH_0^{-1}$ and $D_C$ was defined in Eq. (\ref{eq2}). 

\subsection{Results}

The results of our statistical analyses are shown in Figures 2 and 3. Figures 2a-2c show contours of $1\sigma$, $2\sigma$ and $3\sigma$ on the $\Omega_k -\epsilon$ plane when the Union 2.1 plus SNe Ia SCP-0401, SDSS (SALT2) and SDSS (MLCS2K2) compilations are considered, respectively. For the Union 2.1 plus SNe Ia SCP-0401+ sample, we find that a perfect transparent ($\epsilon=0$) and flat universe is allowed by the current data at $1\sigma$ level, with $\epsilon=0.017\pm 0.055$ ($1\sigma$). For the sake of comparison, we also show the influence of the light-curve fitting on the analysis in Panels 2b and 2c. Note that no significant conflict between them is found, with $\epsilon=0.047 \pm 0.057$ (SALT2) and  $\epsilon=0.067 \pm 0.071$ (MLCS2K2) at 68.3\% (C.L.). { In Figure 3, painels a, b and c display} the likelihood for the $\epsilon$ parameter for Union 2.1 plus SNe Ia SCP-0401, SDSS (SALT2) and SDSS (MLCS2K2) compilations, respectively. In this case we obtain $\epsilon=0.017\pm 0.052$, $\epsilon=0.047\
pm 0.039$  and $\epsilon=0.067\pm 0.056$ at $1\sigma$ level. It is worth mentioning that these 
bounds on $\epsilon$ are only marginally compatible with those found by Avgoustidis et al. (2010), in which a preferably negative value for the opacity parameter was found, i.e., $\epsilon=-0.040^{+0.08}_{-0.07}$ at $2\sigma$ level.

\section{Conclusions}
\label{sec:conclusions}

The results of observational cosmology in the last years have opened up an unprecedented opportunity to test the veracity of a number of cosmological theories as well as the existence of new physics in the Universe. Motivated by these results, several analyses have recently discussed the importance of comparing cosmological distances to explore deviations from the standard cosmological model or a possible presence of exotic physics. However, most of these analyses are cosmological model-dependent, which makes the results less general than desired.

{ In this paper, we have used recent $H(z)$ measurements from passively evolving galaxies to obtain cosmological model-independent distance modulus and impose constraints on cosmic opacity by comparing these data with current SNe Ia observations. In order to perform our analysis, we have considered two  recent samples of SNe Ia, namely, the Union 2.1 + SNe Ia SCP-0401  and the SDSS compilations. We have found that the Union 2.1 plus SNe Ia SCP-0401  compilations is in full agreement with a perfect transparent and flat universe whereas the SDSS compilations are compatible with such a possibility  at $\sim 1\sigma$ level. By marginalizing over the curvature parameter $\Omega_k$, we have found that a completely transparent universe is in agreement with the largest sample in our analysis (Union 2.1 plus SNe Ia SCP-0401). For SDSS sample a such universe it is compatible at  $< 1.5\sigma$ level  regardless the SNe Ia light-curve fitting used.}

\begin{acknowledgments}

The authors thank CNPq and FAPERJ for the grants under which this work was carried out.
\end{acknowledgments}

\end{document}